\documentclass[]{spie} 
\usepackage[utf8]{inputenc}
\usepackage{amsmath,amssymb,bm,commath}
\usepackage{fullpage}
\usepackage{graphicx,tikz}
\usepackage{amsfonts}%
\usepackage{algorithmic,algorithm}
\usepackage{color}
\usepackage{xspace}
\usepackage{comment}
\usepackage[labelfont=bf]{caption}

\newcommand{\xmath}[1]{\ensuremath{#1}\xspace}

\newcommand{\param}[1]{\xmath{\left({#1}\right)}}

\newcommand{\dict}      {\xmath{\bm{\Phi}}}

\newcommand{\datam}     {\xmath{\bm{Y}}}

\newcommand{\coefm}     {\xmath{\bm{A}}}

\definecolor{blue}{rgb}{0.2,0.5,0.7}
\definecolor{green}{rgb}{0.3,0.68,0.29}
\definecolor{purple}{rgb}{0.6,0.31,0.64}

\usepackage[colorlinks=true, allcolors=blue]{hyperref}

\title{Data Processing of Functional Optical Microscopy \\for Neuroscience}

\author[a]{Hadas Benisty}
\author[b]{Alexander Song}
\author[c]{Gal Mishne}
\author[d,*]{Adam S. Charles}
\affil[a]{Yale Neuroscience, New Haven, CT 06510 USA}
\affil[b]{Max Planck Institute for Intelligent Systems, Stuttgart, Germany}
\affil[c]{Halıcıoğlu Data Science Institute, Department of Electrical and Computer Engineering and the Neurosciences Graduate Program, UC San Diego, 9500 Gilman Drive, La Jolla, CA 92093 USA}
\affil[d]{Department of Biomedical Engineering, Kavli Neuroscience Discovery Institute, Center for Imaging Science, Department of Neuroscience, and Mathematical Institute for Data Science, Johns Hopkins University, Baltimore, MD 21287 USA}

\authorinfo{Further author information: (Send correspondence to A.S.C.)\\A.S.C.: E-mail: adamsc@jhu.edu\\  G.M. was supported by funding from the NIH grant no. R01 EB026936.}

\begin{document}

\maketitle

\begin{abstract}
Functional optical imaging in neuroscience is rapidly growing with the development of new optical systems and fluorescence indicators. To realize the potential of these massive spatiotemporal datasets for relating neuronal activity to behavior and stimuli and uncovering local circuits in the brain, accurate automated processing is increasingly essential. In this review, we cover recent computational developments in the full data processing pipeline of functional optical microscopy for neuroscience data and discuss ongoing and emerging challenges. 
\end{abstract}


{\noindent \footnotesize\textbf{*}Adam S. Charles,  {adamsc@jhu.edu}}

\section{Introduction}
Modern neuroscience has been propelled forward by the development of new technologies that offer unique windows into the brain's activity.
These new techniques, including high-density electrodes~\cite{jun2017fully,steinmetz2021neuropixels}, functional ultrasound imaging~\cite{mace2013functional,urban2015real,takahashi2021social}, high magnetic field functional Magnetic Resonance Imaging (fMRI)~\cite{tik2018ultra, viessmann2021high,huber2018ultra} and optical imaging~\cite{bouchard2015swept,demas2021volumetric} all stand to further our fundamental understanding of the brain and new potentials for future therapies. One of the fastest growing segments in this push for advanced neural recording technologies is functional optical microscopy, in particular \emph{in-vivo} recordings using fluorescence microscopy. 

In functional optical imaging of neurons a fluorescing indicator (typically a protein) sensitive to a biomarker of neural activity is introduced to a cell. Example biomarkers include voltage, calcium, potassium, glutamate and sodium etc.~\cite{badura2014fast,kannan2018fast,shen2019genetically,chen2013ultrasensitive,dana2018high}. Out of these, calcium indicators have been the most widespread. When the level of a given biomarker changes (i.e., during or right after a neural firing event), a fluorescent property (e.g. brightness or emission color) of the indicator changes as well. At each image frame the tissue is illuminated with light at a specific wavelength, and any fluorescing indicator have a probability of raising their energy level. When the energy level falls back to the lower-energy state, light at a longer wavelength is emitted and is collected by the microscope. The measured value of the collected light thus reflects the value of the biomarker at a given location. 

Practically, the tissue is illuminated in a number of ways. For example in single photon widefield microscopy, an entire plane is illuminated at once, using a camera to collect full frames simultaneously. This imaging method can acquire high resolution videos at kilohertz framerates, however it is limited in their ability to image deeper regions in highly scattering tissues. Specifically, scattering of light in brain tissue greatly blurs images, requiring optical methods that better localize fluorescence at depth.  Multiphoton imaging, e.g., two- and three- photon microscopy can penetrate deeper tissue $400-1000$~$\mu$m, but rely on raster scanning technologies to iteratively measure small volumes sequentially throughout the plane of imaging~\cite{RN2,jung2004vivo,rodriguez2018three}. 
Hence multiphoton imaging is often used to image smaller structures, like axons, dendrites, and somas, while one-photon imaging is used at meso- and cortex-wide scales~\cite{RN2,sofroniew2016large,yasuda2006supersensitive,homma2009wide} or at somatic resolution under challenging optical conditions, such as via endoscopes and miniscopes in freely moving animals~\cite{jacob2018compact,zhang2019miniscope}.

Fluorescence microscopy holds a number of unique advantages over other imaging methods. Unlike fMRI and functional ultrasound, the activity measured is more directly related to neural activity and not as spatially and temporally blurred by the hemodynamic response~\cite{ma2016wide,nunez2021neural}. Furthermore, optical methods do not require inserting probes into the brain, making them typically less invasive than electrophysiology. The drawback to this advantage is the limited penetration depth of the photons. Imaging deeper structures does require more invasive methods such as implanting a GRIN lens~\cite{xie2006grin,wang2013characterization}.
Additionally, fluorescence microscopy provides entire images that capture not just the neural activity, as electrophysiology can do on a faster time-scale, but also the morphology of the cells. Thus fields of view can be registered across days to enable chronic long-term recordings of the same identified neurons, e.g., during learning. These data, however, are higher-dimensional, and spatiotemporally richer than typical electrical recordings, requiring significant additional computational resources to extract information from.
Moreover, fluorescence microscopy is actively growing and new volumetric imaging techniques ~\cite{bouchard2015swept,demas2021volumetric,song2017volumetric,beaulieu2018simultaneous,demas2021volumetric} promise to even further increase the scale of such data and the spatiotemporal statistics that must be leveraged in the analysis. 
To solve this ``big data" explosion and to process the ever-growing datasets, there is an ongoing need to meet the challenge of designing robust automated algorithms to accurately extract information from these rich data.

Here we discuss the emerging line of work work that has focused on the task of building new analysis tools to realize the full potential of high-resolution large scale imaging. 
The primary goal is extracting time-courses of all the individual units (e.g., neurons, dendrites, brain regions, etc.) from the data, so as to relate these to behavior and stimuli in downstream analyses. This is often accomplished by decomposing the movies into spatial profiles, representing the area in the field-of-view (FOV) that a unit occupies, and the corresponding temporal fluorescence traces.
While this goal is simply stated, the unique properties of \emph{in-vivo} fluorescence imaging create a number of challenges in misalignment of data, tissue aberrations and signal distortion, imaging-dependent noise levels, and severe lack of large-scale ground truth data for validation. These challenges have led to a myriad of approaches, from solving one specific step in the process, e.g., denoising~\cite{lecoq2021removing}, to whole pipeline implementations~\cite{cantu2020ezcalcium}. Approaches also range from imaging of specific structures, e.g., widefield~\cite{saxena2020localized} to methods aimed at many imaging classes~\cite{charles2021graft}. We aim to provide here a walkthrough of the basic challenges, the landscape of current approaches, and finally emerging challenges with no current solutions.

\begin{figure}[t]
    \centering
    \includegraphics[width=\textwidth]{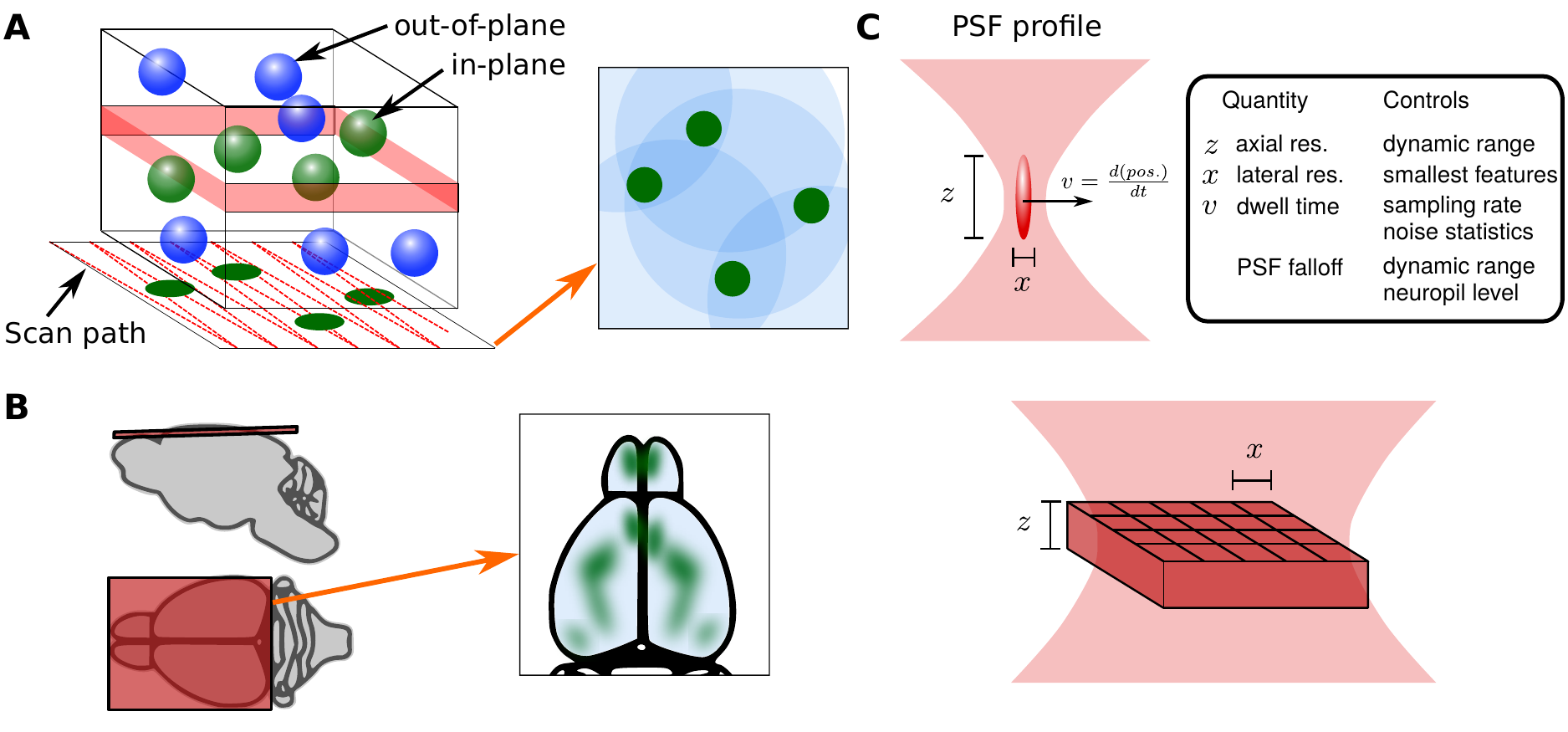}
    \caption{{\bf Optical imaging basics.} A: Laser scanning microscopy (e.g., two-photon microscopy) focus the laser to a point inside the tissue, sequentially illuminating a chosen plane. Neurons intersecting the plane show up in the rendered images, while out-of-plane neurons accumulate as background, or neuropil.
    B: Widefield imaging simultaneously illuminates the entire cortical surface plane, capturing dynamics across multiple brain areas. 
    C: Choices in the optical illumination impact image properties. The image resolution is defined by the point-spread function (PSF), which is determined by microscope optics and optical properties of the sample. In laser scanning microscopy, these, in combination with scanning and laser technology, the image size, resolution, framerate and noise characteristics of the image are determined. In widefield imaging, camera characteristics primarily determine the image acquisition size and rate. The optical resolution for widefield imaging is strongly limited by microscope optics and optical aberrations from the sample.
    }
    \label{fig:intro}
\end{figure}

\section{Functional fluorescence microscopy}
Functional fluorescence microscopy in neuroscience is used to capture the dynamics of neural activity in a wide variety of animal models and targets. Improvements in bioengineering have allowed researchers to study targets from nanometer sized targets in individually labeled neurons to spanning multiple brain regions across centimeter wide fields of view. Recordings are taken at several different timescales and framerates, up to hours long recordings at kilohertz rates. Accordingly, specialized microscopes have been developed to tackle these individual experimental requirements.

Most microscopes can be described by two properties: the illumination and sampling, of which either one or both may be time varying. The illumination is a structured light source that aims to excite fluorophores (indicators) in the targeted region of interest without needlessly exciting other areas. The sampling is a mapping of the light emitted from targeted regions onto the sensor.
The two most common configurations used for functional imaging today are two-photon laser scanning microscopy (TPM) \cite{stosiek2003vivo} (Fig.~\ref{fig:intro}A,C) and widefield microscopy (Fig.~\ref{fig:intro}B,C). 

In TPM the illumination is a focused, pulsed near-infrared (NIR) laser that is raster-scanned in a 2D pattern (Fig.~\ref{fig:intro}A) and the sensor is a single pixel detector (typically a photo-multiplier tube) collecting all emitted light from the whole FOV. The fluorophores are excited via two-photon absorption \cite{zipfel2003nonlinear}, which has a squared-law relationship with the intensity of the excitation beam. This results in the absorption process being confined to a small, point-like focal volume (Fig.~\ref{fig:intro}C) dubbed the point-spread function (PSF). TPM has been widely adopted primarily due to its excellent optical sectioning properties and spatial resolution. Because scattering of NIR light is much lower in brain tissue than visible light and scattering of the fluorescence does not impact image quality, TPM achieves a practical imaging depth of up to 500$\mu m$, even in the highly scattering mammalian or avian brain.

In widefield microscopy (Fig.~\ref{fig:intro}B), the illumination is a visible LED or laser that excites the imaged volume and the sensor is a camera that maps onto the target volume. The excitation light propagates through the whole imaging volume, strongest near the surface of the sample and is attenuated over a couple hundred microns into the sample via tissue absorption. The emitted light is imaged onto the camera, with the contribution from the focal plane mapped most closely onto the sensor and out-of-focus light diffusely collected as well (Fig.~\ref{fig:intro}C). The imaging process is affected by scattering in the emission. In reasonably transparent animal models (such as \emph{C. elegans} or larval zebrafish) this effect is minimal. In highly scattering systems, like the rodent brain, this limits the overall spatial resolution and depth that may be used with this technique to typically $200\mu m$, although smaller contributions from deeper tissue is still present. While the emitted light frequently comes from the whole volume, as in densely labeled samples, genetic targeting or microscopy techniques can isolate the signal to a smaller population of cells. A major advantage of widefield microscopy is the ease of setup in a variety of models and scalability to very large areas and framerates.

Several microscopes have been designed to optimize functional neural recordings. For TPM, development has focused on improving the acquisition speed and volume size. The illumination shape of the PSF has been transformed for imaging sparse tissues effectively \cite{lu2017video} or for volumetric imaging \cite{song2017volumetric}. Three-photon microscopy enables even deeper imaging \cite{ouzounov2017vivo}. Changing the illumination timecourse with custom scanning paths can make fast jumps from cell to cell \cite{katona2012fast}. Live updating of the both scanning pattern and timecourse is used for highs-speed image acquisition \cite{kazemipour2019kilohertz}. For collection, cameras have been used to replace the single pixel detectors to improve acquisition rates \cite{zhang2019kilohertz}. Recently, the combination of multiple techniques and technologies has enabled cellular resolution of up to a million neurons simultaneously \cite{demas2021high}.

Changes to the techniques of widefield microscopy have focused on optical sectioning and volume scanning. Optical sectioning can be achieved by structuring the illumination such that only the plane mapping onto the camera is excited. This has been achieved primarily through light-sheet microscopy, which has been scanned to image 3D volumes in larval zebrafish \cite{vladimirov2014light} and also with a single objective in mice \cite{bouchard2015swept}. Light field microscopy has been used to structure mapping of the sampling from the imaged region to the camera \cite{skocek2018high} and in combination with confocal microscopy enables high-quality optical sectioning \cite{zhang2021imaging}. One of the most major developments has been the miniaturization of microscopes for head-mounted functional imaging in animals, which has greatly expanded the types of problems and brain areas researchers can explore \cite{ghosh2011miniaturized,cai2016shared,scott2018imaging}.

\begin{figure}[t]
    \centering
    \includegraphics[width=\textwidth]{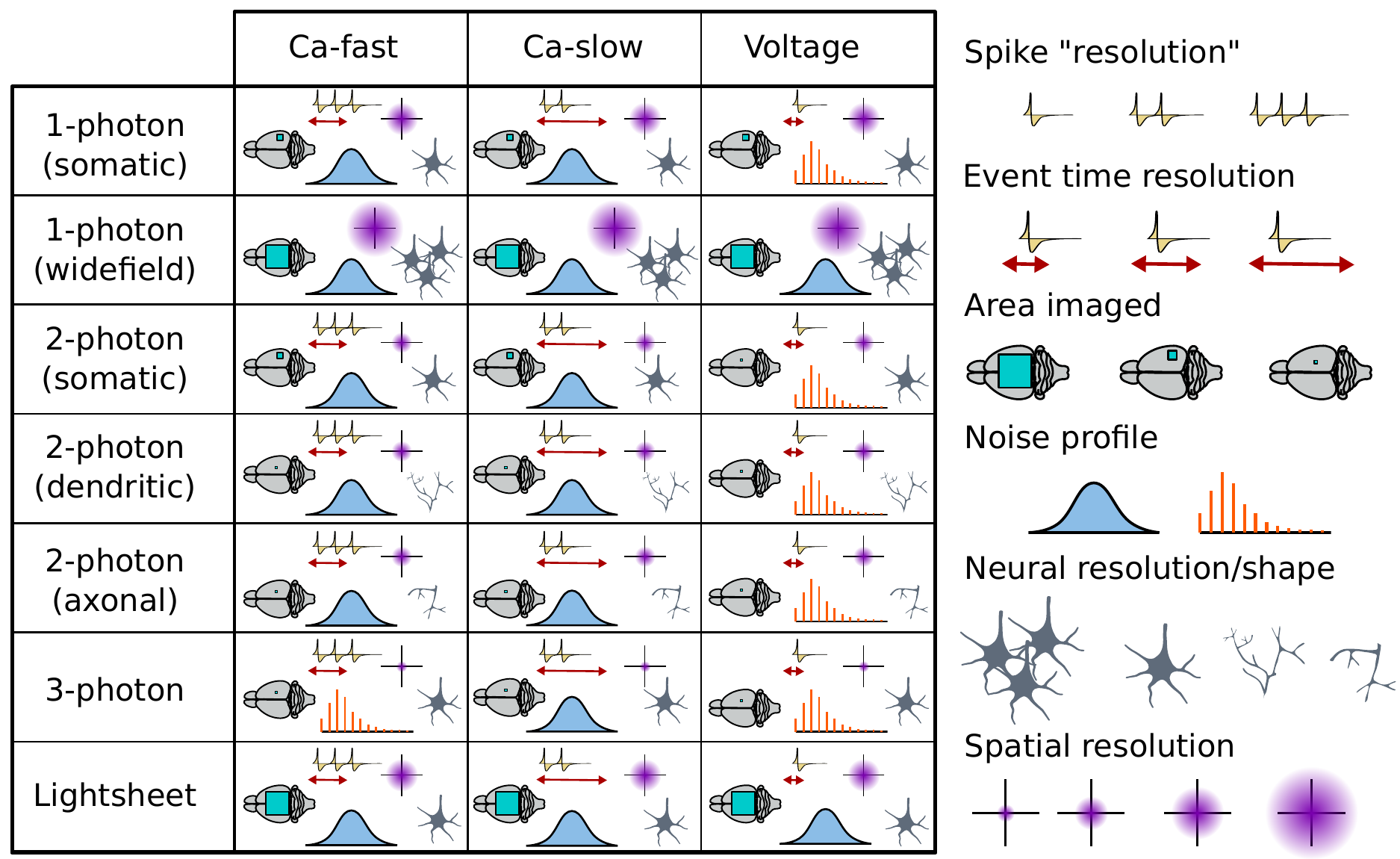}
    \caption{{\bf Imaging domains.} The combination of different indicator and optical properties changes the baseline requirements of the analysis. Fast calcium indicators tend to have more Gaussian-like noise profiles, however produce challenges in identifying single spikes and localizing spikes in time. The effect is more pronounced with slow indicators. Voltage indicators have better temporal statistics but suffer from reduced spatial coverage. Different optics, e.g., 1- 2- and 3- photon imaging also affect the scale, resolution and neural structures imaged.}
    \label{fig:table}
\end{figure}

All of the advances outlined have enriched the space of acquired brain recordings, especially when intersected with the range of available indicators. Specifically, they create a range of spatiotemporal resolutions, signal quality, and imaged morphology, a subsection of which we depict in Figure~\ref{fig:table}. To bridge the gap between the raw data and scientific discovery, these data must be analyzed to extract useful neuronal activity.

\section{Fluorescence Microscopy in space and time}
The current state of functional fluorescence microscopy analysis can be understood through the long history of developments that has led to today's state-of-the-art. Initially, fluorescence microscopy had long scan times and was thus focused on imaging static tissue (see~\cite{ellinger1940fluorescence} for a review of the very early history of fluorescence microscopy for biology). Thus, all the relevant information was anatomical in nature and could be traced manually or, later on, identified automatically to discover complex biological structures. Fluorescence microscopy continues to play an vital role in imaging static anatomical structures and morphological fitting algorithms have evolved with the technology~\cite{kayasandik2016improved,korfhage2020detection,xu2021automated}.

As scan times improved and the ability to inject fluorescent indicators into live tissue emerged, fluorescence microscopy expanded to the imaging of time-varying biomarkers, e.g., voltage indicators in the 1970s~\cite{grinvald1985real} and calcium indicators in the 1980s~\cite{tsien1980new}. With this shift came the added dimension of time, as now fluorescence was a temporal quantity. In neuroscience, this enabled the study of the activity of the neurons over time. Initially, however, labeling methods were in their nascent stage and the use of viruses to introduce fluorescing proteins into cells created a high variability in labeling. The result was often sparsely labeled tissue where overlapping neurons were rare. Without overlap, isolating a single cell's activity could thus be accomplished by identifying individual neurons anatomical extent in the video (e.g., based on the temporal mean or variance of the imaging stack), and then averaging the pixels belonging to each neuron to extract their time-trace. 

To see this, we consider the statistical assumption that each neuron is modulated in a linear way (i.e., the spatial shape and temporal fluctuations are independent). Each $N$-pixel frame $\bm{y}_t\in\mathbb{R}^N$ at time $t$, stacked as an $N\times 1$ vector, can be thought of as the linear combination of the $d$ cell shapes $\bm{a}_i\in\mathbb{R}^N$ for $i=1,...,d$ and its activation at time $t$ $\phi_{it}$:
\begin{gather}
    \bm{y}_t \approx \sum_{i=1}^d \bm{a}_i^T\phi_{it} = \bm{A} \bm{\phi}_t.
\end{gather}
Solving a least-squares optimization of the form
\begin{gather}
    \widehat{\bm{\phi}}_t =  \arg\min_{\bm{\phi}} \|\bm{y}_t - \bm{A} \bm{\phi}\|_2^2
\end{gather}
for the activity of all cells $\widehat{\bm{\phi}}_t$ at time $t$  then can be written as the pseudo-inverse $\widehat{\bm{\phi}}_t = (\bm{A}^T\bm{A})^{-1}\bm{A}^T\bm{y}_t$. With no overlap the Gramm matrix $\bm{A}^T\bm{A}$ is simply a diagonal matrix with the norms of the cell shapes on the diagonal. Thus, the activity for the $i^{th}$ cell at time $t$ is simply a weighted average of the pixel values at that time-point, which has been exactly the methodology with hand-drawn spatial profiles, sometimes also called regions of interest (ROIs). 

As labeling methods advanced and indicator designs provided brighter fluorescent tags, overlaps between neurons and with other processes (e.g., dendrites) became  more common. The increased cell count vastly increased the burden of manual annotation and the increased overlaps mathematically results in a non-diagonal Gramm matrix $\bm{A}^T\bm{A}$, removing the validity of direct averaging. 
Thus, new conceptual approaches were required for cell identification. Rather then solely relying on spatial cues to isolate neurons, activity had to be demixed in space \emph{and} time. 
In fact, as opposed to purely anatomical studies, many modern systems neuroscience analyses abstract away from space, analyzing a neuron-$\times$-time matrix irrelevant of anatomy. Methods, such as dimensionality reduction~\cite{cunningham2014dimensionality,mishne2016hierarchical,benisty2021rapid}, dynamical systems analysis~\cite{vyas2020computation}, statistical connectivity~\cite{pillow2008spatio}, etc. operate on the time-traces and thus suffer more from inexact estimation of the neural activity than the spatial shapes. 

Under the statistical assumption that each neuron is modulated in a linear way, i.e., the spatial profile is constant over time with brightness controlled by a time-vector representing the cell's activity, the problem of isolating neurons can be considered as a matrix factorization problem. If the data matrix $\datam = [\bm{y}_1,...,\bm{y}_T]\in\mathbb{R}^{N\times T}$ is the pixel-by-time fluorescence video matrix, each neuron contributes one rank-1 component 
\begin{gather}
    \datam\approx \sum_{i=1}^d \bm{a}_i\bm{\phi}_i^T = \coefm\dict^T,
\end{gather}
where $\bm{a}_i\in\mathbb{R}^N$ is again the $i^{th}$ neuron's spatial profile (how it appears visually in the data), and $\bm{\phi}_i\in\mathbb{R}^T$ is the $i^{th}$ neuron's time-trace (how the biomarker-driven fluorescence changes over time). 
To find the rank-$d$ decomposition, Mukamel et al. used a combination of PCA and ICA: first PCA was used to reduce the overall dimension of the data and to obtain an initial guess $\datam\approx \sum_{i=1}^d \bm{u}_i\bm{v}_i^T$, and then ICA was performed on the set of vectors $\bm{z}_i = [\alpha \bm{u}_i^T, (1-\alpha)\bm{v}_i^T]^T $. The ICA step  rotates the PCA components into their independent components, demixing both the spatial and the temporal components at once. This procedure allowed for overlapping components and included temporal independence when identifying neurons. 

The matrix factorization approach continued to expand~\cite{pnevmatikakis2016simultaneous,haeffele2019structured,pachitariu2016suite2p}, primarily formulated as an optimization program:
\begin{gather}
    \{\widehat{\coefm}, \widehat{\dict} \} = \arg\min_{\coefm,\dict} \|\datam - \coefm\dict^T\|_F^2 + \mathcal{R}_{\coefm}(\coefm) + \mathcal{R}_{\dict}(\dict), \label{eqn:fullOpt}
\end{gather}
where $\mathcal{R}_{\coefm}(\coefm)$ and $\mathcal{R}_{\dict}(\dict)$ represent appropriate regularization terms over space and time, respectively, that can vary between methods, and often include terms such as the component norms, number of components, sparsity, non-negativity, and spatial cohesion. As direct optimization is often difficult for problems of this size, alternating descent type algorithms are often employed, i.e., iteratively solving 
\begin{gather}
    \widehat{\coefm} | \widehat{\dict}  = \arg\min_{\coefm} \|\datam - \coefm\widehat{\dict}^T\|_F^2 + \mathcal{R}_{\coefm}(\coefm), \nonumber
\\
    \widehat{\dict} | \widehat{\coefm}  = \arg\min_{\dict} \|\datam - \widehat{\coefm}\dict^T\|_F^2 + \mathcal{R}_{\dict}(\dict). \nonumber
\end{gather}
Positive aspects of this approach is that each of these optimization problems can be solved reasonably efficiently (conditioned on judicious choices of regularization) and that different assumptions on the time-traces and component shapes can be incorporated naturally via regularization. 

Thus, in the current landscape of algorithms that extract neural signals from fluorescence microscopy data we now see a combination of approaches. One approach focuses primarily on anatomical identification, leaving the identification of functional traces as a later stage, and another which places identification of the time-traces on equal ground (or more) with the spatial maps and extracts the two jointly. 

\subsection{Methods focusing on space}

The class of methods that focus on anatomical identification have been mostly inspired by image segmentation, using both classical and modern approaches.
Each of these methods relies on a spatial model of the data, either preset or learned from the data. 

One such approach, dictionary learning, stems from the broader image processing literature~\cite{elad2010role}. Dictionary learning assumes a sparse generative model for image patches~\cite{olshausen1996emergence}, and has been applied to calcium imaging in order to learn spatial dictionaries whose atoms represent neuronal shapes. The identified neuron shapes can then be used  to estimate corresponding time courses~\cite{pachitariu2013extracting,Diego2014,SCALPEL}. 
These applications include spatial generative models based on convolutional sparse block coding~\cite{pachitariu2013extracting},  
extensions of convolutional sparse coding to video data with non-uniform and temporally varying background components~\cite{Diego2014}, and 
the dictionary learning of spatial components via iterative merging and clustering~\cite{SCALPEL}. 
 
An alternative approach is to view cell detection as an image segmentation or clustering problem. For examples, in HNCcorr~\cite{spaen2019hnccorr} 
after selecting a set of seeds (superpixels), this method aims to find a cluster within a patch containing the seed by solving the HNC (Hochbaum’s Normalized Cut) problem, which is similar to Normalized Cut. 
The HNC formulation balances a coherence term, which maximizes the total similarity of the pixels within each cluster, with a distinctness term that minimizes the similarity between the cluster and all others. 
Local Selective Spectral Clustering (LSSC)~\cite{mishne2018automated} also solves the cell identification problem as clustering pixels in a high-dimensional feature space. 
Following the construction of a sparse graph whose nodes are the pixels and whose weights are determined by pairwise similarity of the time-traces of the pixels, the nodes are embedded using the eigenvectors of the random-walk graph Laplacian. Pixels are then clustered together in an iterative approach based on selecting subsets of the eigenvectors that best separate a cluster from the rest of the data, and enabling overlapping clusters. After detecting neuronal components (the graph construction allows for morphology-agnostic clustering), time-traces are demixed and calculated by projecting to a low-rank space.  

Another approach based on image segmentation is ABLE (Activity-Based Level set segmentation)~\cite{Reynolds2017}.
ABLE defines multiple coupled active contours~\cite{chan1999active} 
in the field-of-view (FOV) where an active contour seeks to partition a local region into an interior corresponding to a neuronal component and a local exterior, such that the pixels within each region are similar to one another temporally. 
ABLE handles overlapping cells by coupling the evolution of active contours that are close to another.
The evolution of the contours is performed by the level set method~\cite{osher1988fronts}, where only neighboring
cells directly affect a cell’s evolution.
ABLE automatically merges two cells if they are close and temporally correlated, and prunes cells if their size is too small or too large. 
An advantage of this method is that it makes no assumption on a cell’s morphology or temporal activity, therefore it can potentially generalize to different indicators and spatial morphologies.

Modern deep learning methods for image segmentation have also been adapted to the ROI extraction problem, and as opposed to the unsupervised approaches above, are supervised approaches that require training data. 
UNet2DS~\cite{klibisz2017fast} is a neural network based on the fully convolutional UNet~\cite{ronneberger2015u} model, an architecture which was developed for biomedical image segmentation.  UNet2DS takes the mean image as an input and outputs two probability masks of a pixel belonging to either a cell or the background. 
Since U-Net2DS ignores temporal information, it has difficulty separating overlapping neurons and detecting sparsely firing cells.
Such methods also cannot differentiate active cells, i.e., cells that exhibit deviations from baseline fluorescence consistent with spiking events, from non-active cells.
Finally, DISCo (Deep learning, Instance Segmentation, and Correlations)~\cite{kirschbaum2020disco} uses a combination of time-trace correlation-based pixel segmentation on a graph and a convolutional neural network to identify individual spatial profiles. 

\subsection{Methods focusing on space and time}

Rather than leave time-trace estimation as a secondary step, spatiotemporal demixing methods aim to simultaneously identify the spatial maps with their corresponding fluorescence traces. 
Many algorithms thus aim to optimize a cost of the form in~\eqref{eqn:fullOpt}. In particular, given the development of many of these methods on calcium imaging data, they perform matrix factorization with non-negative constraints representing the expected positive deviations from baseline: non-negative matrix factorization (NMF)~\cite{pnevmatikakis2016simultaneous,pachitariu2016suite2p,charles2021graft,haeffele2019structured,inan2017robust,maruyama2014detecting,mishne2019learning}.
The main difference between these methods is the specific choices of regularization and the specific implementation of the algorithmic steps. For example, multiple methods follow the aforementioned alternating optimization~\cite{pnevmatikakis2016simultaneous,pachitariu2016suite2p,charles2021graft,inan2017robust}, while another approach instead uses semidefinite programming~\cite{haeffele2019structured}. Versions of matrix factorization methods have been perhaps the most widespread, with special versions being designed for one-photon endoscopy data~\cite{tran2020automated}, widefield data~\cite{saxena2020localized,charles2021graft}, and voltage data~\cite{buchanan2018penalized}.

Regularization has been applied both on the temporal and spatial dimensions. Temporally, sparsity over the time-traces has been a popular constraint. Specifically, neurons often fire infrequently, relative to the number of frames in a video~\cite{pnevmatikakis2016simultaneous,song2017volumetric}. Taking advantage of this sparsity requires accounting for the biophysically-induced exponential decay of fluorescence. While more sophisticated methods have been developed for post-processing deconvolution (see Sec.~\ref{sec:deconv}), a simple way that has been applied to cheaply deconvolve data is based on the observation that for exponentially decaying responses (i.e., a single-pole response function) with decay rate $\alpha$ can be expressed as the difference equation $F(t) = \alpha F(t-1) + S(t)$, where $S(t)$ is the instantaneous impulses over time. Thus simply computing $\tilde{S}(t) = F(t) - \alpha F(t-1)$ should give a reasonable estimate of a sparse impulse train. The parameter $\alpha$ can further be tuned to maximize sparsity $\max \|S(t)\|_0$. 

Spatial regularization is more complex and often requires a model for neural processes of interest. Initial ideas included ensuring connected components, or by minimizing the total variation (TV) norms to minimize complexity of shapes and remove spurious pixels~\cite{haeffele2019structured}. In more recent versions of matrix factorization algorithms, deep networks trained on human annotated labeled data have been used to classify true and false cells, the resulting system serving as a post-hoc filter that effectively double-checks the spatial regularization. 

As an alternative to shape-based regularization, recent work has leveraged sparsity and spatial correlations in a new way~\cite{charles2021graft}. Instead of using spatial locations of pixels to constrain the components' spatial profiles, Graph-Filtered Temporal (GraFT) dictionary learning builds a graph among all pixels that enables like-pixels to share time-trace decompositions. This approach effectively gives up completely on space, and instead focuses on the learning of a dictionary of time-traces over the data-driven pixel graph.

More recently another important modification to the base optimization in Equation~\eqref{eqn:fullOpt} has been emerging. Rather than focusing on changing the regularization terms used, the least-squares form of the data fidelity term has instead been reconsidered. The least-squares cost enforces a linear-Gaussian data generation hypothesis, however a number of nonlinearities in fluorescence dynamics, incompleteness in component discovery and the non-Gaussian statistics of the photo-diodes all contribute to various extents to errors in demixing. Four main alternatives have emerged, including a robust-statistical approach leveraging a Huber cost function~\cite{inan2017robust}, a contamination aware generative model approach~\cite{gauthier2018detecting}, a zero-Gamma mixture model~\cite{wei2020zero}, and a deep-learning approach~\cite{denis2020deepcinac}. 

Convolutional deep learning networks taking into account temporal statistics or activity have have also been developed~\cite{apthorpe2016automatic,soltanian2019fast,bao2021segmentation}.
A (2+1)D convolutional neural network~\cite{apthorpe2016automatic} was trained on spatiotemporal sliding window and the output represented the probability of a pixel belonging to an ROI centroid.  
Apthorpe et al. demonstrated that adding the temporal domain helps suppress noisy detections compared to a 2D network that only took as input the time-averaged image.
However, the network they trained only learned spatial 2D kernels. In comparison, STNeuroNet~\cite{soltanian2019fast} is a 3D convolutional neural network based on DenseVNet (similar to a U-Net) also trained on overlapping spatiotemporal blocks, and its output is a binarized probability map. 
By adding a temporal max-pooling layer to a typical DenseVNet architecture, the spatiotemporal features are reduced to a spatial output, which increases the speed of training and inference. This gain is important for network validation and low-latency inference in closed-loop experiments.
Shallow U-Net Neuron Segmentation (SUNS)~\cite{bao2021segmentation} aims to simplify the neural network architecture of STNeuroNet to further improve speed, while still incorporating temporal information. To this end, a temporal matched filter tailored for shot noise is applied to the input video, thereby enhancing calcium transients occurring across multiple frames, and reducing temporal information into individual 2D frames.
Preprocessing also includes a whitening process, which normalizes the fluorescence time series of each pixel by the estimated noise of that pixel. This yields an SNR representation that highlighted active neurons and obscured inactive neurons.
This SNR video is then processed by a shallow UNet, whose output is a probability map. 
Post-processing for both methods includes clustering of the output probability maps to detect individual cells and aggregation of detected cells across the probability maps for merging and pruning.
Note that while these methods identify cells based on a spatiotemporal analysis, they do not address the issue of estimating time-traces.

\section{Imaging analysis pipeline}
\begin{figure}[t!]
    \centering
    \includegraphics[width=0.7\textwidth]{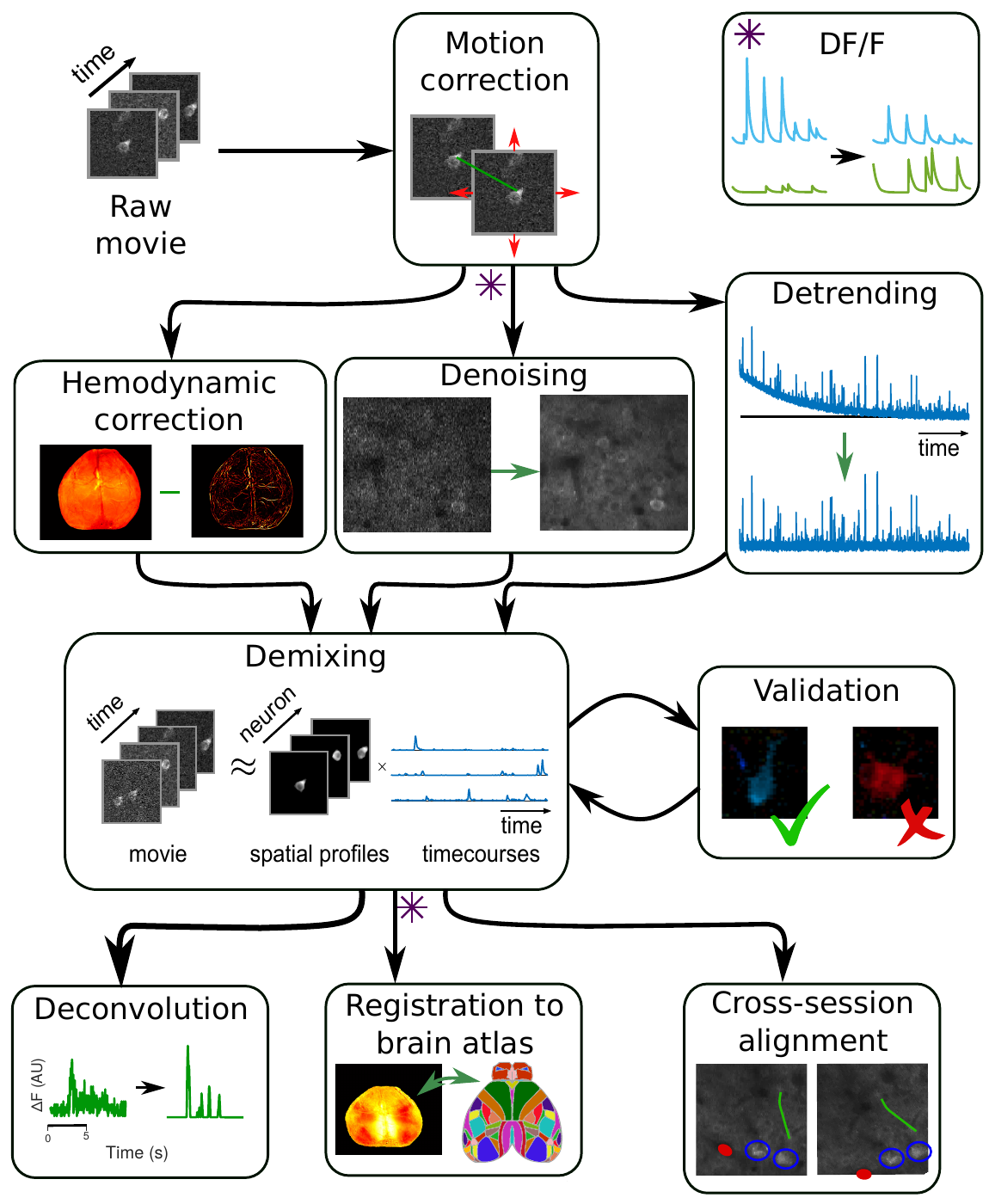}
    \caption{{\bf The optical imaging pipeline, expanded.} The first step in analyzing fluorescence microscopy data is to correct any motion, reducing inter-frame misalignment. Between motion correction and demixing (i.e., cell-finding), steps to remove imaging noise and confounding factors are taken. Specific steps include de-trending to remove photobleaching, denoising to increase SNR, and hemodynamic correction in widefield data. The core pipeline stage of focus is demixing, or identification of individual fluorescing components. Demixing should always be paired with some sort of validation on the output to prevent artifacts and other errors. The output of the demixing can be used to infer firing events (in single-cell resolution imaging), align multiple imaging sessions, or register to a global brain atlas (in widefield data). One important step: the DF/F calculation which normalized fluorescence traces to their baseline, can be computed before or after demixing. }
    \label{fig:pipeline}
\end{figure}

The fundamental output of an optical functional imaging analysis pipeline are the identified spatial profiles, and more importantly their corresponding time-traces. While demixing thus serves as the core of analysis, often multiple preprocessing and post-processing steps are typically part of the pipeline to facilitate this output (Fig.~\ref{fig:pipeline}).  
Motion correction is typically the first step in the analysis pipeline, to register all frames in the imaging stack such that the neuronal components to be extracted occupy the same spatial footprint in all frames. 
Denoising can be applied as a preprocessing step either temporally~\cite{charles2021graft} or spatially~\cite{mishne2018automated} to improve the detection of ROIs.
Normalizing per-pixel time-traces, e.g., by z-scoring, can enhance dim cells, and improve cell detection. 
Following ROI extraction, post-processing in the temporal domain can include the following: neuropil estimation, denoising of time-traces (if not implicitly part of the extraction itself) and calculating normalized time-traces ($\Delta F / F$).
Post-processing in the spatial domain can include automatic classification~\cite{giovannucci2018caiman} of identified components into true or false neuronal components or manual quality control. 
To identify single spiking events, deconvolution is a post-processing step that aims to recover sparse firing events from the fluorescence time-traces.
Finally, in longitudinal studies, it will also include registration of imaging sessions and matching of ROIs across sessions.

\subsection{Motion correction}
The majority of ROI extraction methods rely on neuronal components being within a fixed/consistent spatial footprint in the FOV - thus necessitating image registration of the individual frames.
Motion between frames can be due to several factors~\cite{stringer2019computational,laffray2011adaptive,chen2013online,collman2010high,sekiguchi2016imaging}: animal motion during imaging (e.g., locomotion), scanning artifacts, mechanical strain, drift relative to the objective, changes in the brain, e.g., due to hydration, etc.
Given the length of imaging sessions (tens of thousands of time frames), computational complexity of the registration approach is an important consideration.  
For fast acquisition rates inter-frame motion can be considered as a global constant offset of the FOV, therefore rigid registration of translational shifts is sufficient~\cite{guizar2008efficient,dubbs2016moco,mitani2018real}. 
For example, registration can be performed to a reference (template) image using cross-correlation or phase-correlation~\cite{pachitariu2016suite2p}. 
The reference image is typically set as an average over the initial frames and then regularly updated as an average over later subsets of frames, or an average over the full stack. The reference frame can be made more precise by an iterative refinement procedure to reduce blurring~\cite{pachitariu2016suite2p,hattori2021patchwarp}.
As an alternative to performing correlation-based translation registration, bright cells can be detected and tracked over time using particle tracking~\cite{aghayee2017particle}. 
A rigid (translation and rotation) transform is then calculated for each frame to the next by minimizing the residual displacements of all tracked cells. This approach can also be applied to volumetric imaging. 

Non-rigid motion artifacts can occur due to lower acquisition speeds, which result in artifacts such as shearing in later scanned lines of the image, or slow distortions over long recording sessions due to mechanical (z-drift with respect to the objective) and biological issues (warping of brain tissue due to metabolic activity, dilation of blood vessels, and liquid reward consumption)~\cite{pachitariu2018drift}.
Non rigid motion correction usually relies on splitting the image into overlapping spatial patches and performing registration at the patch level. This registration can be rigid at sub-pixel resolution~\cite{pnevmatikakis2017normcorre,pachitariu2016suite2p} or a more flexible affine transform~\cite{hattori2021patchwarp}.

Most methods for motion correction target two-photon somatic imaging. 
However, such methods can struggle with non-somatic neuronal components such as dendrites, due to the difference in size and impact of z-drift.
While a cell-body is typically on the order of $\approx$15~$\mu$m---with variation depending on brain area, species etc.---the width of an axon or a dendrite is $\approx$1 micron. Thus, slight registration errors can have a significant effect on identifying the spatial footprint of these components. 
Furthermore, z-drift can cause segments of tuft dendrites to shift in and out of the field-of-view.
This can lead to difficulties in aligning the dendrite to the reference image. From a computer vision perspective, this can be thought of as image registration under occlusions. 

Another potential complication is the use of GRIN lenses to image deeper structures in the brain. Optical aberrations near the edges of GRIN lenses can significantly change the motion characteristics, requiring distortion-aware realignment. 
Finally, niche technologies can also create novel situations, such as the rotating platform developed for near-freely moving imaging~\cite{voigts2018animal,voigts2020somatic}.

\subsection{Denoising and normalization}

The absolute noise levels present in optical imaging creates a challenging signal extraction environment. It is only because many sequential frames of the same population are recorded that individual cells can be identified. This process, however, can be improved by modest noise filtering as a preprocessing stage. A number of methods (with example applications) are used across the literature, including median or low-pass filtering~\cite{song2017volumetric,mishne2018automated}, downsampling, PCA projection~\cite{pachitariu2016suite2p,mishne2018automated}, z-scoring, wavelet denoising~\cite{charles2021graft}, other hierarchical models~\cite{charles2017stochastic}, and deep learning-based denoising~\cite{lecoq2021removing,li2021reinforcing}. All these approaches make different noise and signal model assumptions and should be used judiciously. 

For example, median/low-pass filtering and downsampling, are simple, quick steps that can be run on the data at early stages. Median filtering is effective at reducing shot noise common in low-photon environments, at the cost of making shapes in the image more ``convex'' (i.e., filling in corners). Low-pass filtering and downsampling reduce high-frequency noise, however blur the data via the convolutional filter. Downsampling~\cite{friedrich2017multi,apthorpe2016automatic} has the further benefit of reducing the data size in space or time, thus reducing later computational costs, however effectively reduces the sampling resolution. 

Other signal models are more complex. Hierarchical models can provide flexible ways of both incorporating different noise classes (e.g., Poisson) and flexible signal models (e.g., via inter-pixel correlations), however at a heavy computational overhead~\cite{charles2017stochastic}. Wavelet-based denoising~\cite{donoho1994ideal,chang2000adaptive} is perhaps the most versatile in this class, as both per-image and per-pixel time-trace denoising are computationally efficient, readily implemented across programming languages, and can handle sharp transitions, thereby reducing blurring.

Yet other methods model the noise in ways based on the data itself. PCA-based projections identify a low-dimensional space that captures much of the data variance, effectively assuming that low-variance principal components represent noise.
Sparsely firing cells or dim cells, however, can often end up in low-variance components and thus removed with the noise. 
Penalized matrix decomposition~\cite{buchanan2018penalized} denoises the imaging data by performing a patch-wise penalized low-rank decomposition.  
Recently, general advances from deep learning~\cite{lecoq2021removing,li2021reinforcing,weigert2018content} have been proposed for denoising calcium imaging.

For example, DeepInterpolation~\cite{lecoq2021removing} uses a clever design that trains a neural network to predict a movie frame based on previous and past frames. Independent noise, which cannot be predicted, is thus filtered out. 
A broader image-restoration method CARE~\cite{wei2020comparison}, uses a U-net trained on high- and low-resolution images to enhance signal quality. 
As with all deep-learning methods, the drawbacks include training the network (if the pre-trained options do not fit the application), and a general lack of knowledge as to the exact expected biases of the black box system. 

Usually performed at the same time as denoising, normalization plays an important role in numerical stability of algorithms. Having data with large (or small) overall fluorescence values can cause problems in the condition numbers of the matrices used in optimizing costs in the demixing step. Normalizing, e.g., to unit-median or unit-max values helps constrain these effects, and all normalization can be undone after demixing to return meaningful fluorescence values to the identified time-traces. 
As an example, methods based on deep networks typically employ preprocessing to ensure appropriate dynamic range across the data, regardless of background or other inhomogeneities in the illumination (e.g., via homomorphic filtering~\cite{bao2021segmentation}).  

As a final note, a form of normalization that often is required in preprocessing is de-trending to remove photobleaching~\cite{nolan2018algorithms} (see~\cite{kubler2021robust,cutrale2019using} for examples). Photobleaching is the reduction in overall fluorescence over time stemming from the fluorescent proteins becoming trapped in an intermediate quantum state and becoming inactive. The result is a large imbalance over time in the dynamic range of the signal, which can hinder most demixing algorithms. 

\subsection{Neuropil estimation}
In 2p imaging, neuropil is a ``background" signal that contain the fluorescence of neuronal elements that are out-of-focus (e.g., dendritic, axonal) and scattering. 
This signal can contaminate the estimated fluorescence of an ROI if it is not properly accounted for.
Some methods estimate neuropil as part of the ROI extraction process by automatically identifying the signal of the exterior surrounding the cell~\cite{Reynolds2017}, or explicitly or implicitly adding at least one background signal in the linear decomposition model~\cite{pnevmatikakis2016simultaneous,giovannucci2018caiman,charles2021graft,mishne2019learning}, for example
\begin{gather}
    \datam\approx \sum_{i=1}^N \bm{a}_i\bm{\phi}_i^T+\bm{a}_{\textrm{b}}\phi_{\textrm{b}}^T 
\end{gather}
where $\bm{a}_{\textrm{b}}$ and $\phi_{\textrm{b}}$ are the spatial and temporal components of the background,  respectively. 
In other methods, neuropil is estimated in a post-hoc process by calculating the average time-trace from the pixels within the ring surrounding each extracted spatial profile~\cite{pachitariu2016suite2p,keemink2018fissa}.
The signal is then subtracted from the time-trace of the spatial profile.

\subsection{Normalized fluorescence ($\Delta$F/F)}

Neurons have different concentrations of fluorescence indicator, e.g., GCaMP, which can result in varying levels of fluorescence both across cell populations and between individual neurons \cite{dana2014thy1,daigle2018suite}. Additional variation also occurs across the image field of view due to imaging technique, microscope optics, and sample variation.

Therefore, to obtain comparable signals from varying neural sources, extracted time-traces are normalized \cite{helmchen2011calibration} to remove the baseline fluorescence activity and adjust the amplitude as follows:
\begin{equation}
    \frac{\Delta F}{F}(t) = \frac{F(t)-F_0}{F_0},
    \label{eq:dfoverf}
\end{equation}
where $F(t)$ is the extracted time-trace and $F_0$ is the baseline fluorescence.

Calculation of baseline fluorescence relies on the assumption that neurons have sparse activity, thus a baseline level that does not correspond to actual neuronal activity can be estimated using varying heuristics. A common method of estimating $F_0$ is averaging the activity of the time frames with $p=10\%$ lowest activity
(the exact value of $p$ can vary from lab to lab, for widefield imaging $p=50\%$ may be used to instead describe the percent change from the mean activity level). This method assumes neurons are quiescent for long enough periods that the change in fluorescence level is effectively zero at this percentile. A lower percentile is not used to account for error caused by sources such as shot noise or axial motion.

The running percentile calculation additionally assumes there is little to no fluorescence from other sources. In datasets that are densely labeled or have high background fluorescence, the technique will consistently underestimate the true dF/F as contributions from other sources gets added to the baseline estimate. Estimating the baseline fluorescence with overlapping sources requires separating the fluorescence contributions from individual sources and any background source, which is automatically achieved in most segmentation algorithms \cite{pnevmatikakis2016simultaneous,pachitariu2016suite2p}. This approach may be more reliable than the running percentile calculation by integrating additional information provided by activity transients. A test using synthetic data demonstrated these estimates of baseline fluorescence were reliable for the brightest cells and another estimation procedure making use of pixelwise spatial information further improved estimates of baseline activity \cite{song2021neural}.

\subsection{Deconvolution} 
\label{sec:deconv}
The product of the core demixing stage includes a set of time-traces---one per component---that contain the temporal fluorescence fluctuations. Due to the complex interactions and resulting buffering of non-voltage fluorescence-inducing biomarkers, the transient increases in fluorescence due to individual spiking events is stretched out over time. For example, calcium indicators can observe deviations from baseline in the recorded fluorescence from a single spiking even for seconds.

As spiking events are typically considered the primary source of neural communication, efforts to infer underlying spiking from fluorescent data have emerged, taking a number of forms~\cite{Diego2014,pachitariu2018robustness,friedrich2017fast,evans2019use,jewell2020fast,pnevmatikakis2013sparse,shibue2020deconvolution,theis2016benchmarking,rupprecht2021database}.
Core to all is the attempt to invert a generative model of fluorescence $F(t)$ as a function of spike-times $\{\tau_1, \tau_2, ..., \tau_K\}$. 
The full biophysical model dictates that at each time $\tau_k$ of a spike event, a stochastic influx of the biomarker (e.g., calcium) flows into the cell and drives a  nonlinear differential equations and nonlinearity that determine the level of bound fluorescent protein over time. 
While the full biophysical model includes nonlinear differential equations and nonlinearities~\cite{helmchen2015single,lutcke2013inference,song2021neural}, this model can be linearly approximated as 
\begin{gather}
    F(t) = \sum_{k=1}^K g_k h(t-\tau_k) + e(t)  
\end{gather}
where $h(t)$ is the response curve (e.g., exponential rise-and-decay) to a single event and $g_k$ represents the stochasticity of the influx of biomarker at the $k^{th}$ event. 

Given the convolutional form of this model, identifying the time-points of events has taken the form of deconvolution. These algorithms have taken the form of assuming functional forms over $h(\cdot)$, such as exponential~\cite{pnevmatikakis2016simultaneous} or double exponential~\cite{pachitariu2017robustness} kernels, or taken more model-free approaches in a deep-learning framework~\cite{theis2016benchmarking,rupprecht2021database,speiser2017fast}. 
Specific algorithms have also varied widely, including exact $\ell_0$-regularized optimization~\cite{jewell2020fast}, marked point process~\cite{shibue2020deconvolution}, interior point optimization~\cite{vogelstein2010fast}, active-set methods~\cite{friedrich2016fast}, Variational Autoencoders~\cite{speiser2017fast}. 

One of the core difficulties in deconvolution is the probabilistic relationship between spikes and fluorescence. In addition to the biomarker levels being variable, simultaneous recordings of electrophysiology and calcium imaging show that at times optical imaging can miss single events at significant levels (i.e., missing 70\%-80\% of individual spikes)~\cite{huang2021relationship}. Thus, in general across noise, indicators, and other experimental conditions, deconvolution may have highly varying performance limits.  
In fact, for assessment, benchmarking attempts have avoided direct timing comparisons, opting instead for using local rate averages over $>40$~ms bins as a validation metric~\cite{theis2016benchmarking}. 
Interestingly, spike ambiguity in optical imaging has seeded another approach: to remove spiking events from the equation---literally---by marginalizing out the spiking events and directly estimating a latent spike rate~\cite{ganmor2016direct}.

\subsection{Multi-session registration}
The ability to record from the same \emph{identified} population of neurons with functional optical imaging in longitudinal experiments across multiple days is a major advance and enables understanding long-term processes such as learning and memory.
One of the crucial post-processing steps of such longitudinal experiments is the alignment of the recorded imaging data across days to enable one-to-one mapping of neurons across all sessions. 
This is essential to understanding the changes in neural representation over time.
However, alignment is challenging due to the 3D non-rigid transformations between imaging sessions. These are a result of, for example, day-to-day variance in the imaging angle due to slight changes in the angular placement of the microscope objective, day-to-day variance in optical clarity of cranial windows, and changes in the brain tissue over days. 
Specifically in TPM this can lead to differences in the shape of recorded cells, since slight z-drift and tilts can result in relatively large changes in the cross-section of a cell.
Semi-automated approaches to calculate the transform between imaging sessions and match neurons exist, however these rely on user input to select matching ROIs and only align pairs of sessions~\cite{pachitariu2016suite2p}.
Recent methods propose registering imaging sessions based on fully affine invariant methods originally developed for natural image registration~\cite{li2020fully}.
A recent approach~\cite{yadav2022multi} based on the classical SIFT~\cite{lowe2004distinctive} algorithm enables fast automatic registration of calcium imaging sessions, and one-to-one matching of ROIs, even if the neuron was not detected in all sessions. 

In one-photon imaging alignment is challenging due to light scattering and lack of optical sectioning, which increase
the similarity between the time-traces of neighboring neurons in the FOV~\cite{sheintuch2017tracking}. In addition, only active cells can be tracked, as opposed to multi-photon imaging. 
Sheintuch et al.~\cite{sheintuch2017tracking} developed a  probabilistic method for automated registration across one-photon imaging sessions that is adaptive and optimized to different datasets.
First, all cells are mapped to the same image by registering each session to a reference session using a rigid transformation based on the centroid locations of extracted ROIs. 
Next, the probability for any pair of neighboring cells from different sessions to be the same cell is calculated, given their spatial correlation and centroid distance and a probabilistic model for similar and dissimilar matched cells.
Cells are finally aligned across sessions by an iterative procedure based on the estimated probabilities.

\section{Modern challenges in optical functional imaging}

\subsection{Imaging morphologies beyond the soma}

More recently, variants of optical imaging have aimed to expand the scope of accessible brain signals by imaging both larger and smaller neural structures. At one end of this spectrum, zooming in enables the imaging of dendritic and spine structures, which captures how individual neurons communicate~\cite{denk1996imaging,kerlin2019functional,suratkal2021imaging,graves2021visualizing,ali2019interpreting,sabatini2001ca2+}. Dendritic~\cite{xu2012nonlinear} and axonal~\cite{broussard2018vivo} imaging, while also having sparse temporal statistics, can have long, thin spatial profiles that span the entire field-of-view (FOV). We note that these types of neural morphologies are also vital in some species, such as \emph{Drosophila melanogaster}, where dendritic activity is vital to tracking neural processing~\cite{seelig2010two,vajente2020calcium}. 
Approaches to dendritic/axonal imaging have largely followed in the path of somatic imaging analysis. For example recent versions of Suite2p~\cite{pachitariu2016suite2p} can be run in ``dendrite mode''. The long, stringy morphologies, however, are at odds with typical built-in assumptions of spatial locality. A more recent approach instead re-defines pixel ordering on a data-driven graph to better identify irregular morphologies~\cite{charles2021graft}. 

At the other end of the spectrum, cortex-wide (i.e., widefield) imaging can be achieved at resolutions too coarse to isolate activity signals of individual neurons, but instead can capture brain-wide activity patterns~\cite{scott2018imaging}.
As widefield data exhibits significantly different  statistics from micron-resolution imaging both in time and space, we provide a more thorough discussion later in Section~\ref{sec:wf}.

\subsection{Voltage imaging}

Voltage imaging is a technique that has been used for decades to record changes in neural activity using voltage-sensitive dyes ~\cite{ebner1995use}. Voltage imaging, as compared to recording changes in calcium, is a more direct way of measuring neural signals. A comparison of population responses to optical recordings using calcium indicators and voltage indicators showed major differences in the temporal response of the recorded calcium signal as compared to the voltage signal~\cite{zhu2021population}. Widefield voltage imaging has also been explored~\cite{mohajerani2010mirrored} and faces many of the same challenges as widefield imaging with calcium sensors (see Section~\ref{sec:wf}).

Technology for optical recording of voltage signals has improved rapidly over the past few years. The development of improved voltage sensors in the form of bright genetically encoded voltage indicators (GEVIs) has enabled high resolution voltage recordings at multiple spatial scales~\cite{knopfel2019optical}. Genetic targeting of these indicators to subcellular structures isolates signal to particular neuronal structures, further increasing the signal-background ratio (SBR). These improvements have allowed researchers to generate optical voltage recordings from a population of cells in awake, behaving animals~\cite{piatkevich2019population}.

Several challenges still exist that prevent the generation of large scale optical voltage recordings. Voltage indicators are membrane-bound, which limits the total concentration of the sensors as compared to calcium indicators, which may fill the whole cytoplasm. Generally, GEVIs are, at least for now, dimmer than their GECIs counterparts. In order to take advantage of the improved temporal response function, recordings must be made at much higher framerates as compared to calcium indicators ($\sim1000 Hz$ vs $\sim30Hz$). The combination of these challenges reduces the overall spatial scale of current cellular or sub-cellular resolution recordings with voltage indicators.

Current voltage imaging analysis includes both matrix factorization~\cite{buchanan2018penalized,xie2021high} and deep learning~\cite{cai2021volpy} demixing approaches similar to methods used in calcium imaging. 
The potential for low-SNR and non-Gaussian noise statistics can complicate demixing. Moreover extremely high temporal resolutions create much larger datasets, increasing the computational cost of processing. 
Finally, non-negativity is a basic assumption built into many calcium imaging analysis methods: i.e., deviations from baseline are only positive. Voltage traces have no such constraint. 

\subsection{Widefield imaging}
\label{sec:wf}
Multi-photon microscopy provides a way to record neuronal activity of individual cells and cell parts from a particular FOV, typically limited to a few hundred microns. These recordings allow to thoroughly investigate local micro-circuits comprised from hundreds to thousands of cells at a time.
Widefield imaging, however, trades cellular resolution for increased FOV (millimeters) and enables exploration of the overall activity through imaging of the entire cortical surface \cite{cardin2020mesoscopic}. Collection of widefield signals is performed using a scientific CMOS camera, capable of imaging hundreds of frames per second. To control the acquired data size and increase framerate, most researchers reduce the spatial resolution to $512\times512$ pixels, resulting in $10^2$-$10^4 \mu$m spatial resolution.

The extracted time-traces reflect an aggregated summary of the neuronal activity captured from thousands of cells, cellular compartments and depths (although mostly superficial layers \cite{ma2016wide}). Estimation of spike rates is usually not performed for widefield signals as the captured signals may originate from various cell parts such as axons, dendrites as well as somas, each related to a different kernel. While this issue can be resolved by calcium indicators targeting specific parts of the cell at the expense of limited temporal resolution \cite{bengtson2010nuclear, kim2014prolonged}, most researchers find the standard GCaMP indicators as the (currently) best in terms of spatial and temporal resolution for capturing synaptic activity, providing a valuable tool for exploration of the dynamics of large-scale networks and their relation to complex behavior, perception and cognition \cite{cramer2019vivo, ren2021characterizing}.

Preprocessing of widefield recordings typically includes 4 stages: alignment, normalization, hemodynamics correction and parcellation. Below we describe the motivation for each stage and common practices. 
Imaging of the entire cortical surface naturally enables analysis/modeling across animals. To facilitate a one-to-one correspondence of data acquired from different animals, frames of each session are registered to align to a global template of the cortex, according to several anatomical control points using an affine transform \cite{musall2019single}.
In many cases the captured time-traces exhibit a slow decrease in baseline activity, ascribed to bleaching. This effect is easily removed by subtracting the slow trend (evaluated by low-pass filtering) from each pixel \cite{barson2020simultaneous}. To equalize spatial differences of expression levels, each pixel is normalized with respect to its own overall variance post detrending. 
 
The following stage of preprocessing aims to correct the hemodynamics artifacts which is unique to widefield signals and is not present in 2-photon imaging. Fluctuations in blood flow and oxygenation alter excitation and emission of photons due to Hemoglobin absorption. This phenomenon contaminates the captured signals with unwanted dynamic components \cite{ma2016wide}. The most common approach to correct this artifact is to alternate the emitted light with an additional reference channel, for example, UV light ($\sim$ 400nm). As GCaMP6 is isosbestic to UV light, the emitted photons are assumed to be independent of neuronal activity whereas the data channel (typically blue $\sim$488nm) will cause emission of photons affected by fluctuations of both neuronal activity and hemodynamics signals. 

The Beer-Lambert law is an exponential model for the measured light intensity as a function of wavelength, absorption and traveled path. Assuming that temporal deviations from the average signal are small (as they often are for widefield calcium imaging) this relation is simplified by taking a first order estimation of the signal in a given pixel $i$:
\begin{equation}
    {y}_{t}\param{i} = \frac{F^{\textrm{blue}}_t\param{i}}{F
    ^{\textrm{UV}}_t\param{i}}\frac{\overline{F}^{\textrm{blue}}\param{i}}{\overline{F}^{\textrm{blue}}\param{i}}
    \label{eq:pixelwise_hemo}
\end{equation}
where $\bm{F}^{\textrm{blue}}$ and $\bm{F}^{\textrm{UV}}$ are $N$-dimensional vectors of the recorded signals at time $t$ through the blue and UV channels respectively and $N$ is the number of pixels in a frame. The corrected signal, $\bm{y}_t$, is evaluated using a pixelwise linear regression \cite{barson2020simultaneous,tian2009imaging}. 
Recently, a computational approach for improving hemodynamics reduction based on a single reference wavelength was proposed \cite{lohani2020dual}. This approach exploits spatial dependencies between pixels by formulating a multivariate model:
\begin{eqnarray}
        \bm{F}^{\textrm{blue}}_{t} &=& \bm{z}_{t} + \bm{y}_{t} + \nu_{t}\nonumber\\
        \bm{F}^{\textrm{UV}}_{t} &=& \bm{A}\bm{z}_{t} + \xi_{t}
\end{eqnarray}
where $\bm{F}^{\textrm{blue}},\bm{F}^{\textrm{UV}},\bm{y},\bm{z},\nu$ and $\xi$ are n-dimensional vectors, and $n$ is the number of pixels included in a certain local patch. The corrected signal is estimated using the optimal linear predictor, $\widehat{\bm{y}} = \bm{H}_1 \bm{F}^{\textrm{blue}} + \bm{H}_2 \bm{F}^{\textrm{UV}}$ where the matrices $\bm{H}_1,\bm{H}_2$ are evaluated from the signals at both wavelengths. Taking the patch size to be 1, this approach reduces to pixelwise regression \eqref{eq:pixelwise_hemo} where using $n>1$ leads to improved reduction of the hemodynamics artifact \cite{lohani2020dual}.
An alternative approach relies on using two reference wavelength (thus alternating three channels altogether) to obtain a more accurate correction of hemodynamic absorption \cite{valley2020separation}.

The extracted signals at this stage are time-traces of the activity at individual pixels. These traces are high dimensional (typically over $10^4$ pixels) and subjected to several noise sources (e.g., electronic, photonic shot noise). The final stage of widefield preprocessing is therefore to extract a compact representation of brain activity and filter out the noise component as much as possible. Most methods for extracting this representation can be formulated as a linear decomposition model:
\begin{equation}
    \bm{Y} \approx \bm{A}\bm{\Phi}^T 
    \label{eq:parcellation_represenation_matrix}
\end{equation}
where $\bm{Y}$ is a $N \times T$ matrix of the activity of $N$ pixels at $T$ time frames, $\bm{A}$ is a $N \times d$ matrix of spatial components, $\bm{\phi}^T$ is a $d \times T$ matrix of temporal components and $d<N$.
Different methods for decomposing $\bm{Y}$ vary from solely relying on anatomical features to being completely data-driven which affects the spatial interpretability of $\bm{A}$ accordingly. Choosing one method over another should be done considering what downstream analyses will be used and the overall biological hypothesis of the research.

Using Singular Value Decomposition (SVD) to reduce the dimension of widefield data relies on the fair assumption that the variance of neuronal activity within the widefield signal is significantly higher than the noise variance. The activity is decomposed to $\bm{Y} = \bm{U} \bm{S} \bm{V}^T$ where $\bm{S}$ is a diagonal matrix of the singular values and $\bm{U}, \bm{V}$ are orthogonal matrices and a low-dimensional representation \eqref{eq:parcellation_represenation_matrix} is obtained by setting:
\begin{eqnarray}
   \bm{\Phi}^T &=& \begin{pmatrix}
                                s_1 &        & \\
                                    & \ddots & \\
                                    &        & s_d
                       \end{pmatrix}
                       \begin{pmatrix}
                                \bm{v}_1 & \hdots & \bm{v}_d      
                        \end{pmatrix}^T\nonumber, \quad \quad
    \bm{A} = \begin{pmatrix}
                                \bm{u}_1 & \hdots & \bm{u}_d      
                        \end{pmatrix}
    \label{eq:svd}
\end{eqnarray}
where $\bm{v}_i$ and $\bm{u}_i$ are the columns of $\bm{V}$ and $\bm{U}$ respectively and $s_i$ are the singular values.
The number of components, $d$, is selected so that at least 80-90\% of the signal can be reconstructed, assuming that the remaining 10-20\% relates to noise. The spatial components $\bm{u}_i$ are not constrained to be localized or non-negative and therefore the temporal components extracted in~\eqref{eq:svd} do not indicate a trace of activity related to a specific brain region. 
Post-processing is performed in the reduced dimension domain and then projected back to the full dimension using $\bm{U}$. For example, in \cite{musall2019single}, SVD components were used to measure how well external variables (behavior, stimuli) can predict brain activity. The trained regression parameters, each computed per temporal component, were projected to the brain-mask domain in order to biologically interpret the statistical findings. 

A different approach is to take advantage of the spatial structure of widefield signals where adjacent pixels are typically highly correlated while noise is uncorrelated. Therefore a compact and (spatially) filtered representation can be obtained by dividing the brain into localized sub-regions (parcels) and extracting the average trace within each parcel. In this case $\bm{A}$ is non-negative where each row relates to a specific brain parcel with the corresponding column in $\bm{\phi}$ as the time-trace of activity.
Unlike imaging of local circuits, where identifying cell boundaries is a well defined task (although not simple for automation), detection of parcels boundaries is not straightforward.

Identification of sub-cortical borders can be performed experimentally by presenting sensory stimuli e.g. the auditory cortex \cite{romero2020cellular} or the visual cortex \cite{sit2020distributed}. These methods are highly efficient for detecting boundaries of sub-regions within a specific cortical area (visual cortex, auditory cortex), but cannot be used to detect regions that are not responsive to sensory stimuli.

The most common approach for cortex-wide parcellation is to use a pre-defined a atlas based on anatomical features such as the Common Coordinate Framework (CCFv3) proposed by the Allen Institute for Brain Research \cite{oh2014mesoscale} or the mouse brain atlas of Paxinos and Franklin \cite{paxinos2001kbj}. Parcellation based on anatomical atlases presents many advantages - each brain parcel represents a well known biological functionality (e.g. vision, motor, sensory) and a straightforward way to compare neuronal activity across animals (and studies). 
However, it is often observed that the spatial patterns of activity in some regions is not well described by anatomical outlines. Therefore, computational methods for functional parcellation has been a major target of research in recent years for calcium imaging as well as in the fMRI community \cite{zhi2021evaluating}. 

Localized semi-nonnegative matrix factorization (LocaNMF) is a recently proposed approach aiming to tackled this issue by formulating an optimization problem for minimizing the mean square error between the widefield signal and the estimated signal in \eqref{eq:parcellation_represenation_matrix} such that the columns of $\bm{A}$ are non-negative and localized according to anatomical clusters \cite{saxena2020localized}. 
LocaNMF produces, by nature, spatial patterns that are similar to the anatomical boundaries used by the optimization process and therefore typically does not deviate much from the anatomical atlas. 
Related to NMF, a linear one-hidden-layer autoencoder has also been used to identify parcellations in auditory cortex~\cite{liu2019parallel}. 
Similar in spirit, GraFT~\cite{charles2021graft}, being agnostic to spatial morphology due to its underlying graph-based modeling, has also been applied to extracting (potentially overlapping) widefield spatial maps in rat and mouse macroscopic data.
In a recent study, functional parcellation took a different turn by adding a temporal component to brain parcellation. This approach is based on finding repeated spatiotemporal patterns of activation termed motifs, that can be viewed as a time-varying brain parcel. The overall activity is therefore represented as a sum of convolution terms between each motif and its corresponding time-trace~\cite{macdowell2020low}. 

A different approach for brain parcellation is to cluster the brain into regions of co-activity, with no regard to anatomical features. In this case, the matrix $\bm{A}$ is comprised from binary vectors, each corresponding to a specific brain parcel. Li et al. proposed an iterative greedy algorithm for parcellating the brain based on correlation similarity \cite{li2019density}. Other approaches use correlations as graph weights connecting pixels serving as nodes. Parcels are obtained by clustering the graph using Ncut, where the number of parcels is a hyper parameter \cite{lake2018spanning}, or a greedy adaptation of spectral clustering, where the number of parcels is learned from the data~\cite{cidan}.

Overall, the product of functional parcellation methods describe the spatial distribution of co-activity in a given session (per animal). These patterns may be consistent across sessions~\cite{cidan} but are not, in general, uniform across animal and obviously vary with the nature of the experiment (e.g. spontaneous activity, task directed). In that regard, as performed for the SVD approach, post-processing values produced per functional parcel (e.g. goodness of fit, modeling coefficients) can be projected onto the brain mask using the spatial components.

Mapping of cortical neuronal activity is also recently addressed experimentally through multi-modal imaging where widefield calcium imaging recording is performed simultaneously with fRMI signals \cite{lake2020simultaneous} or imaging of local circuit using 2-photon microscopy \cite{barson2020simultaneous} or electrophysiology  \cite{xiao2017mapping, clancy2019locomotion,liu2021multimodal, peters2021striatal} where specific regions are identified as highly correlated to a specific cell (or sub-population).

\begin{figure}[t]
    \centering
    \includegraphics[width=0.9\textwidth]{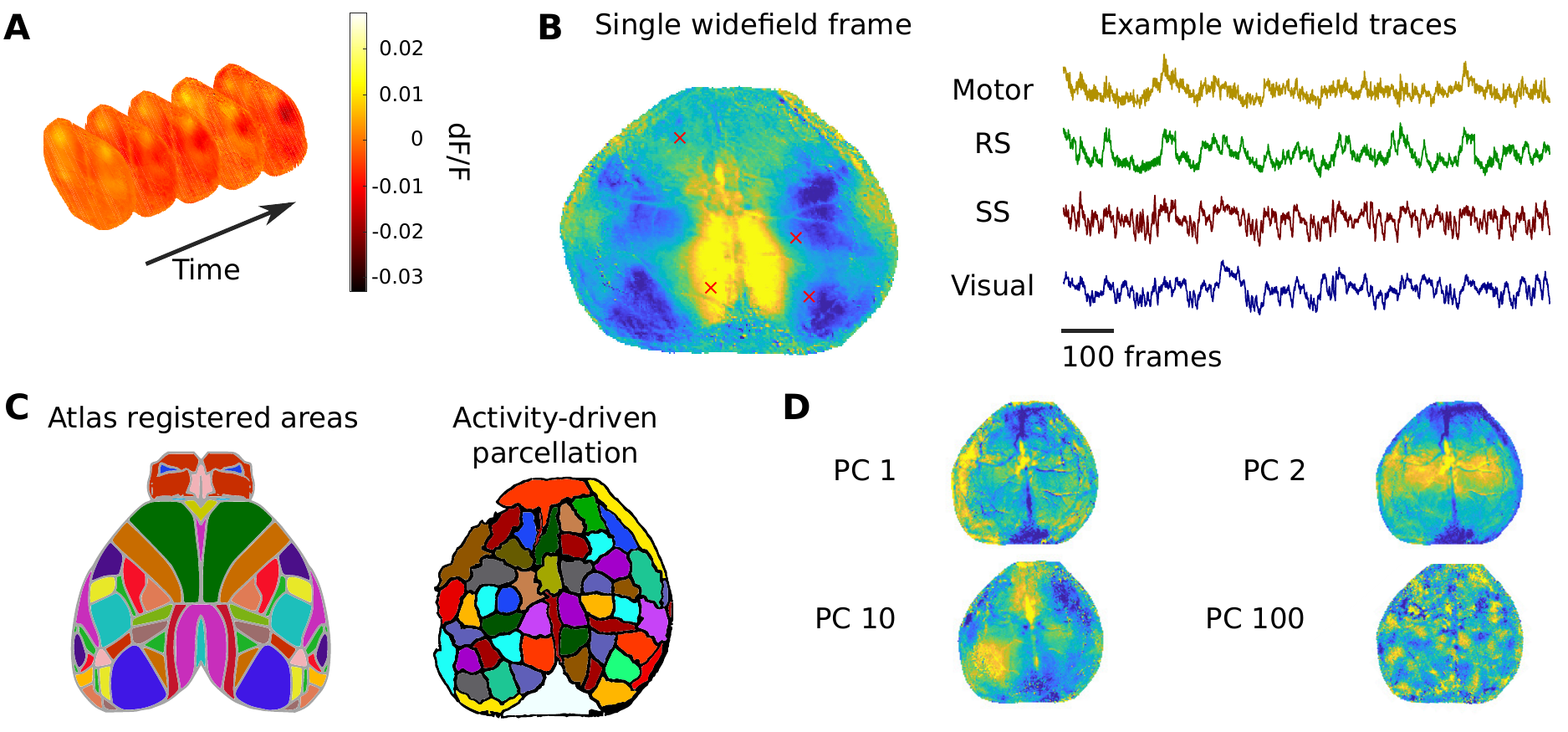}
    \caption{{\bf Widefield imaging.} A: Imaging frames of widefield signal. B: left - a single frame, red 'x' correspond to: Motor, retrosplenial (RS), somatosensory (SS) and Visual cortices, right - time-traces of the corresponding locations. C: left - anatomical atlas CCFv3, right - functional parcellation by LSSC. D - Spatial components derived by SVD. Data from \texttt{http://repository.cshl.edu/id/eprint/38599/}}
    \label{fig:widefield}
\end{figure}

\subsection{Computational imaging}
 
One critical avenue that may completely upend much of how optical imaging data is processed is computational imaging. Computational imaging represents a paradigm wherein optical and algorithmic components are co-designed to compensate and enhance each other and achieve superior results to advances in either area independently~\cite{mait2018computational,waller2020physics}. Co-designed approaches are nascent in \emph{in-vivo} functional imaging of the brain, with only a few examples aimed at faster imaging~\cite{kazemipour2019kilohertz,deb2021programmable} or volumetric imaging~\cite{song2017volumetric}. The algorithmic designs for computational imaging tend to require specialized and often unique processing elements that invert the optical path of the co-designed microscope~\cite{waller2020physics}. Thus, while basic denoising or techniques discussed might still be applicable, more highly coded optics may not even be able to use motion correction, let alone all the advances in demixing. Thus new frontiers are constantly expanding and requiring novel advances in our handling of functional optical data.  

\section{Validation and assessment}

One of the most difficult tasks in creating widely applicable and robust calcium image processing methods is proper assessment~\cite{pnevmatikakis2019analysis}. The effects of mismatch between the necessary simplified statistical assumptions in the signal processing models and the actual data properties must be explored in terms of the effect on the fidelity of extracted signals and, when possible, the later scientific analyses. A prime example of such an effect is explored in~\cite{gauthier2018detecting}, in which it is shown that the \emph{i.i.d.} Gaussian noise assumption often considered can create bleed-through between overlapping cells and additional fluorescent biological processes in the tissue. The result is that the time-traces have high levels (between 15\%-25\%) of transient events being false transients in that they do not reflect activity of that cell. While significant in and of itself, it is further noted that these errors can cause errors in the \emph{interpretation} of the neural activity, including skewed discovery of location encoding in the hippocampus. 

To identify the accuracy of optical imaging processing algorithms, a number of avenues have emerged. Specifically four current available avenues are assessment based on 1) manual annotation, 2) biophysical simulation, 3) local self consistency of global decompositions, and 4) consistency based on external experimental variables. 

We note that a fifth form of validating signals extracted from optical measurements take the form of simultaneous electrophysiological recording and optical imaging~\cite{huang2021relationship,theis2016benchmarking,berens2017standardizing}. The number of simultaneous cells that can be recorded during imaging within a FOV, however, is limited. Thus these recordings have primarily taken a role in assessing either the accuracy of estimated spikes from optically recorded calcium~\cite{theis2016benchmarking,berens2017standardizing}, or the similarity between electrically and optically recorded neural signals~\cite{huang2021relationship}. As yet, this approach does not meet the scale required to completely assess the processing of a full FOV.

\subsection{Manual annotation}
The most basic assessments of optical imaging analysis is the manual labeling of cells. 
This is typically done by annotating an image summarizing the structure in an imaging session, i.e., temporal max projection, temporal average, local temporal correlation~\cite{smith2010parallel}, or nonlinear embeddings~\cite{cheng2020spectral}.
We note that while manual labeling is typically done on the processed data, labels can also be obtained via anatomical imaging (e.g., z-stacks or nuclear labeling with activity-invariant fluorescent proteins). Comparison to manual annotation gives clear metrics: hits (manually identified cells that overlapped significantly with a match in the returned spatial profiles) and misses (those with no match) (Fig.~\ref{fig:assess}A). 

However, there are limitations to assessment via manual annotation. Time-traces are not available as ground truth and must be inferred from the data given the annotations. Additionally, the annotations are often incomplete, excluding sparsely firing cells, or non-somatic components. The effect is that found profiles that do not match well to the manual annotations could appear as `false alarms' but actually fall into many categories. They may be actual cells in the data missed by the annotator, they may be algorithmic artifacts caused by merging or splitting parts of cells~\cite{gauthier2018detecting}, or they may be overfitting to noise or neuropil. Thus at best, manual annotations give a lower bound on true hits and missed detections, but not much information on false positives. 
Some of the limitations stemming from human error can be removed in new datasets consisting of electron microscopy reconstructions of imaged tissue~\cite{zhou2018efficient}. This dataset, and any that follow, provide anatomical ground truth in the form of a registerable volume to match spatial components to, although time-traces are still unavailable.

\subsection{Biophysical simulations}
One of the predominant forms of assessment is \emph{in-silico} simulation~\cite{song2021neural,li2020calciumgan}. Simulation plays an important role in assessing the fundamental limits of algorithms across signal processing and machine learning. In particular, simulations are vital when ground truth information is difficult to obtain, either efficiently, or at all. Functional fluorescence microscopy is exactly one such situation. The time, effort, and expense of simultaneous recording  fluorescence imaging and electrophysiology is a remarkable effort for only a limited portion of the ground-truth data. Simulations offer a potential solution by leveraging anatomical and physical knowledge of the system to generate data where the underlying activity and anatomy driving the synthetic observations are completely known and can be compared to (Fig.~\ref{fig:assess}B).  

Simulations, however, pose their own risk. Simplistic simulations can miss complexities in the real-data imaging statistics, e.g., non-Gaussian or non-\emph{i.i.d.} noise. Complex simulations run the opposite risk of being so detailed that the computational run-time and memory requirements become excessive and capture details irrelevant to the assessment being conducted. Recent work on the Neural Anatomy and Optical Microscopy (NAOMi) simulator seeks to balance these competing needs~\cite{song2021neural}. To date NAOMi has been applied to a number of scenarios, such as testing different demixing algorithms~\cite{charles2021graft}, denoising methods~\cite{lecoq2021removing}, sensitivity to negative transients~\cite{vanwalleghem2021calcium}, and testing/training calcium imaging demixing algorithms~\cite{rupprecht2021database}. 

\subsection{Data consistency}
A third form of assessment uses no ground truth data, manual or synthetic, and instead focuses on self-consistency of the data model. The spatial profiles and time-traces give a global decomposition of the data by minimizing data fidelity and regularization terms over the entire dataset. In this approach, one can focus on a smaller segment of the data and check if the global decomposition matches the local statistics. Specifically, one can check if in the movie frames during which a given cell was purported to have fired (a transient burst in the time-trace during those frames), if the shape of the cell truly appears in the video. Recent work used this local averaging idea to identify how many algorithms are not locally consistent: i.e., activity from different sources  bleed into each other~\cite{gauthier2018detecting} (Fig.~\ref{fig:assess}C). The resulting errors (termed false transients) can influence scientific findings, and can appear in the time-trace estimates of many different algorithms and across many datasets~\cite{gauthier2018detecting}. To address this finding, additional work has sought to develop more robust time-trace estimators that prevent these errors~\cite{inan2017robust,gauthier2018detecting,denis2020deepcinac}.

\subsection{Consistency with external measures}
\begin{figure}[t!]
    \centering
    \includegraphics[width=\textwidth]{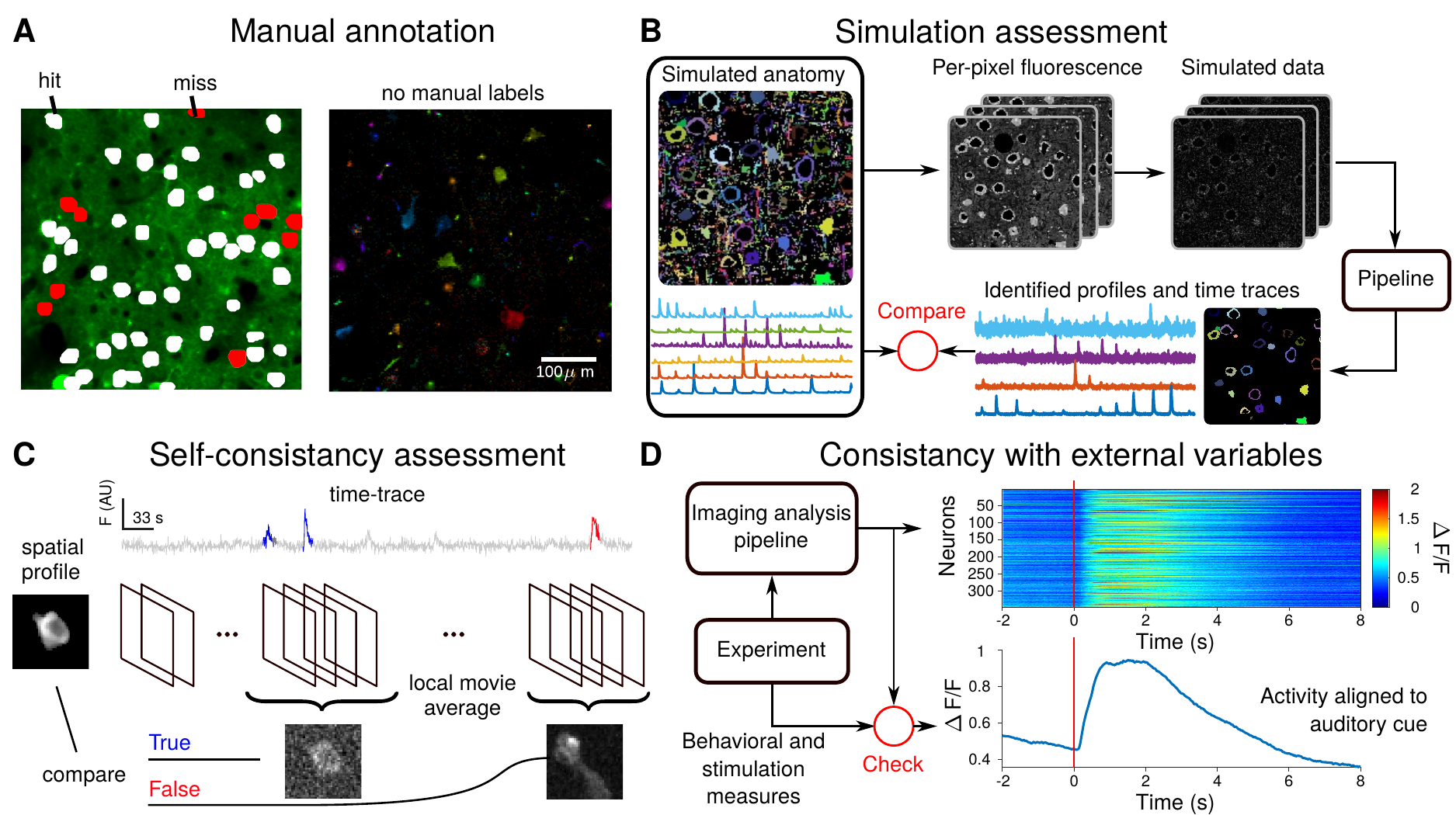}
    \caption{{\bf Assessing optical imaging analysis outputs.} A: Manual annotation, e.g., the NeuroFinder benchmark dataset (example shown here), can help detect true detections and missed cells, as compared to a labeling of the spatial area imaged (right). Manual annotation, however, does not provide information on cells that do not match any of the manually annotated cells (left). 
    B: Simulation-based assessment uses \emph{in silico} simulations of activity and anatomy to create faux fluorescence microscopy data. The outputs of an analysis pipeline using the data can be compared to the simulated ground truth.
    C: Self consistency measures use the global demixed components and analyze local space-time extents of the movie to check if the spatial and temporal components still describe the data well. For example, is activity erroneously attributed to a found cell.
    D: Consistency with external variables examines if expected gross behavior of the imaged neural populations match behavioral or stimulation events co-recorded with the fluorescence video. For example, shown here are neurons in motor cortex responding to a motion made by a mouse responding to an auditory cue.~\cite{levy2020cell}}
    \label{fig:assess}
\end{figure}

The final form of validation we discuss is with respect to external measures outside the imaged brain area. 
To explore the relation between brain activity and behavior, perception and cognition most experimental setups record, simultaneously with the brain activity,  external variables such as spontaneous behavior (whisking, running, pupil size), responses to sensory stimuli and task related behavior. A thorough examination of the extracted traces of activity with respect to external variables can be an important tool for assessment of fluorescence microscopy processing. 
For example, aligning the activity traces to external onsets such as repeated presentations of a sensory cue, specific trained behavior or even running onsets allows to examine brain activity in a behavioral context. Applying this strategy should be done with cautious as uninstructed behavior may cause significant trial-to-trial variability~\cite{musall2019single}. Still, using well pronounced signals such as the response to a visual or auditory cues mostly leads to a significant increase of activity in the visual/auditory cortices~\cite{glickfeld2013mouse, kato2015flexible} or to a distinct pattern of activation of the motor cortex in well trained animals~\cite{levy2020cell} (Fig.~\ref{fig:assess}D). For widefield imaging these same strategies can be applied with respect to the appropriate brain parcels. If no sensory cues are presented, averaging across running onsets should lead to a significant increase of the overall activity of the cortex \cite{lohani2020dual}. 
Alternatively, additional modalities of neuronal activity can be used to for validation of calcium imaging signals such as electrophysiology~\cite{wei2020comparison}, fMRI~\cite{lake2020simultaneous} or dual imaging of widefield and 2-photon imaging~\cite{barson2020simultaneous}.

\section{Discussion}

We have attempted to review here a large portion of the literature related to the analysis of functional optical microscopy data. The topics we have covered aim to provide a practical overview of how optics choices affect signal processing challenges, and the many methods that have been developed to solve these challenges. While broad categories, such as denoising, motion correction, and in particular signal extraction, have been explored in detail in the literature, there are many other challenges that have yet to be solved. 

For one, robust alignment across sessions is an ongoing challenge. While matching cells based on anatomical morphology is currently possible, effects such as nonlinear shearing in the brain and axial drift can create situations where only portions of a field-of-view can be recovered and aligned across recording sessions. Even within long sessions, these effects can remove cells from parts of the recording or reduce the signal quality. Identifying the periods where cells are not visible is critical as the cell is not inactive in these epochs, which may be assumed by subsequent analyses were the time-traces to be padded with zeros. Instead, missing data methods must be employed, which need to know exactly when the missing data occurred. 
As chronic recordings become more commonplace, we expect that cross-session alignment and axial shift compensation will become critical hurdles to pass. 

Another growing area of interest is the real-time analysis of functional microscopy data. Real-time analysis enables closing of the loop, i.e., the ability to use estimates of neural activity to drive future experimental trials~\cite{charles2018dethroning}. While basic manual annotation is trivial to move to an online setting, full demixing algorithms that solve many of the aforementioned challenges are still in their nascent stages. Initial work is promising in the ability to motion correct~\cite{aghayee2017particle} and infer calcium activity~\cite{giovannucci2017onacid,giovannucci2021fiola}. However, the ability to infer activity and cell shapes completely online for dense neural fields is still an ongoing research direction.

A critical aspect of analysis methods that we have not discussed here is the computational cost across methods. There are large variations in cost from simple averaging to training entire deep neural networks. Unfortunately in some regards these disparities match the high variability in the computational infrastructure available across labs. Moreover, as optical imaging of neurons continues to advance, computationally efficient techniques will only become more critical. Already at moderate fields of view there are up to $10^3$ neurons. New methods leveraging the latest microscope designs are recording $10^5-10^6$ neurons at once~\cite{demas2021volumetric,bouchard2015swept}. Furthermore, volumetric (3D) imaging complicates~\cite{Botcherby2006,lu2016,RN3,RN14,Yang2016,RN5,RN15,song2017volumetric} image analysis by rendering many planar segmentation methods unapplicable. In these cases new methods will need to be developed, and computational efficiency will be key to analyzing these very high dimensional datasets. 

Another topic we have not elaborated upon is the effect of indicators on signal analysis. We have focused primarily on generic properties of fast- and slow- calcium indicators and, as a faster comparison, voltage indicators. 
Most indicators share basic characteristics with these classes. In the spirit of universal analysis pipelines, one possible approach is to further abstract all indicators into basic quantities, e.g., quantum efficiency, coupling affinity, etc. which can be used to tune parameters in more generic versions of the current algorithms. This approach, which requires precise estimate of these quantities and their variation, would greatly benefit the user base. 

Another change with different indicators is the assumption of non-negativity. Calcium imaging analysis has largely assumed only positive deviations from baseline. Voltage imaging and widefield calcium imaging all display negative dips as well, which will represent more degrees of changes that need to be implemented into more general pipelines. We note that even for calcium imaging, interestingly new explorations discuss the presence of so-called negative transients, as well as the ability of various algorithms to cope with these unexpected dips~\cite{vanwalleghem2021calcium}.

One increasingly popular group of approaches in the analysis of functional optical imaging is to leverage advances in deep learning. Critical to the success of deep learning approaches is 1) the availability of training data and 2) the generalization of the trained system to new datasets. 
For training data, most systems use NeuroFinder~\cite{berens2017standardizing} and/or the Allen Institute Mouse Brain Observatory~\cite{Allen2019url}. However, both datasets suffer from both mislabeled data and missing labels, or erroneous time-trace estimates~\cite{gauthier2018detecting}.
Generalization is tougher to ensure, as image statistics can affect deep learning systems in a myriad of unexpected ways. Therefore extensive experimentation is required, for example by testing a trained system on imaging from different depths to explore the effect of tissue distortion~\cite{soltanian2019fast}. 
Thus, despite the fact that pre-trained networks are fast to run, these challenges create a high level of uncertainty in using pre-trained networks accurately. 
Individual labs can instead choose to annotate their own data to train networks from scratch, such that they are optimized for the imaging used locally. While this solves some of the aforementioned challenges (assuming care is taken in annotation), the training procedure can be computationally intensive relative to a lab's computational capabilities.

One path to improving training data has been to augment the dataset, such as using random rotations and reflections~\cite{mikolajczyk2018data}. For optical imaging, augmentation should use the degrees of variation known through characterizations of optical-tissue interactions. The use of biophysical simulators to generate synthetic training data, or even modulate real data, can prove useful. In fact recent work used the NOAMi simulation suite~\cite{song2021neural} to generate data for training a spike estimation network~\cite{rupprecht2021database}. 
Finally, another consideration in deep learning approaches is a heavy class imbalance both spatially and temporally between neurons and background in functional fluorescence imaging datasets, i.e. background frequently dominates the acquired FOV, and the desired neural activity may be temporally sparse. 

As we have noted, assessment of data segmentation quality is, in general, challenging.
Evaluation datasets must reflect typical use cases of the community, and therefore they need to constantly be updated and expanded. Due to the rapid expansion of functional microscopy technology, the typical use case is already eclipsing standard datasets, requiring the description of new benchmarks. New datasets are beginning to fill the void, e.g., in voltage imaging~\cite{cai2021volpy}. However, the quality and diversity of imaging data is only accelerating. For example, it would be immensely valuable to the community to provide benchmark data for especially difficult situations, such as cortex-wide widefield data.
Even more powerful would be an amalgamation of different benchmark datasets---both real and synthetic---similar to what SpikeForest~\cite{magland2020spikeforest} has done for spike sorting algorithms. Only with such a wide-ranging set of tests will the strengths and weaknesses of different algorithms become apparent.

Assessment is even more difficult in widefield data, where anatomy provides minimal aid in assessing the validity of spatial component shapes. Neighboring brain areas can share tracts of coordinated activity, and activity can likewise be constrained to a small portion of a brain area. Parcellations thus need to be carefully considered in the context of the behavior and other recordings. In this line of consideration is the nature of parcellations themselves. An unsolved question currently facing the community is whether parcels can overlap: current methods include both solutions that allow~\cite{charles2021graft, liu2019parallel} or prohibit~\cite{li2019density,lake2018spanning,cidan} overlapping parcels. Both confer different interpretations: non overlapping parcels provide a solution wherein each component represents a given brain area's activity, while overlapping can enable the time-traces to be better event- or behavior-locked in the case of a brain region having multiple uses. Future work should explore the interplay between these two windows into widefield data.  

One final major concern is the reproducibility and accessibility of functional imaging analysis techniques (which we note is not restricted to this modality~\cite{charles2020toward}). 
Making code available and easier to use~\cite{cantu2020ezcalcium,romano2017integrated} is only the first step in this path. Learning the intricacies in robustly running the software in new hardware environments, such as local clusters or desktops, is a hard-earned expertise. One option is to ensure that the software is robust and tested on many systems. This approach requires either systems engineering skills that are not typically within the budget or scope of a typical research lab. Instead, a concerted effort across an entire community is required. The community has to coalesce around---and contribute actively to---a specific approach. An alternative being explored is to have individual labs containerize their software, i.e., create shareable virtual environments that are tested with the given software. In neuroscience, an emerging example is the NeuroCAAS system~\cite{abe2021neuroscience} which provides a Dockerized implementations of algorithms to be run, e.g., on the amazon web service (AWS). 
 
In conclusion, functional optical imaging in neuroscience is rapidly growing and accurate, automated processing of the massive data being generated is becoming increasingly essential to the continued progress of understanding the brain. Solving these challenges will take both the forms of new methods that enable, e.g., real-time, robust, fast analyses, and well engineered infrastructure that democratizes the current advances to the growing number of labs employing functional microscopy in their experiments. 
We thus expect that this area will continue to grow rapidly in the next decade, drawing on increased interest from labs across neuroscience, data science, imaging, and other related disciplines. 


\end{document}